\documentclass[unsortedaddress,prd]{revtex4-1}

\usepackage{amsbsy}
\usepackage{amssymb}
\usepackage{amsmath}
\usepackage{geometry}
\usepackage[british]{babel}
\usepackage{amscd}
\usepackage{fancyhdr}
\usepackage{hyperref}
\usepackage{mathrsfs}
\usepackage[dvips]{graphicx}

	\newcommand{\ncd}{\newcommand}
	\ncd{\mrm}    {\mathrm}
	\ncd{\beq} {\begin{equation}}
	\ncd{\eeq} {\end{equation}}

	\def\d{{\rm d}}

\begin{document}

	\title{Conformally invariant thermodynamics of a Maxwell-Dilaton black hole.}
	\author{C S Lopez-Monsalvo}
	\email{cesar.slm@correo.nucleares.unam.mx}
	\affiliation{Instituto de Ciencias Nucleares\\
     			 Universidad Nacional Aut\'onoma de M\'exico, A. P. 70-543, M\'exico D. F. 04510, M\'exico}

	\author{F Nettel}
	\email{fnettel@ciencias.unam.mx}
	\affiliation{Departamento de F\'\i sica, Fac. de Ciencias \\
			 Universidad Nacional Aut\'onoma de M\'exico, A. P. 50-542, M\'exico  D. F. 04510, M\'exico}
	
	\author{H Quevedo}
	\email{quevedo@nucleares.unam.mx}
	\affiliation{Instituto de Ciencias Nucleares\\
     			 Universidad Nacional Aut\'onoma de M\'exico, A. P. 70-543, M\'exico D. F. 04510, M\'exico}

	\date{\today}

	\begin{abstract}
	The thermodynamics of  Maxwell-Dilaton (dirty) black holes has been extensively studied. It has served as a fertile ground to test ideas about temperature through various definitions of surface gravity. In this paper, we make an independent analysis of this black hole solution in both, Einstein and Jordan, frames. We explore a set of definitions for the surface gravity and observe the different predictions they make for the near extremal configuration of this black hole.  Finally, motivated by the singularity structure in the interior of the event horizon, we use a holographic argument to remove the micro-states from the disconnected region of this solution. In this manner, we construct a frame independent entropy from which we obtain a temperature which agrees with the standard results in the non-extremal regime, and  has a desirable behaviour around the extremal configurations according to the third law of black hole mechanics.  

	\end{abstract}

\maketitle

\section{Introduction}

 The link between thermodynamics and relativity has opened up new roads in our quest to probe gravity at a more fundamental level. Over the past thirty years a lot of work has been made to understand the connection between the laws of thermodynamics and gravitational phenomena \cite{hawking77, bardeen, bekenstein}. This is possible through the identification of the surface gravity at the horizon of a black hole spacetime with the thermodynamic temperature of such a system \cite{hawking77}. One can safely use this argument in the case of stationary -  and even quasi-stationary - spacetimes, in agreement with the local equilibrium hypothesis of equilibrium thermodynamics. It is commonly accepted that black holes fullfil the third law of black hole mechanics \cite{bardeen}, i.e. when the extremal configuration is reached the entropy as well the temperature must vanish. Although, it has been proposed that the extremal configuration may represent a critical point where a phase transition occurs \cite{cai}, it is not clear how to assing an entropy to a naked singularity. Then, one can adopt this third law as a useful criterion to test some definitions for the entropy and temperature of black holes.

In this paper we address the puzzle posed by Garfinkle et-al \cite{ghs} concerning the invariance of the temperature of a charged black hole of string theory when working in two conformally related frames. In their article, they found the Hawking temperature of the solution through Euclidean techniques. The striking feature is that the temperature does not depend on the charge of the black hole. Thus, it is not sensitive to the near extremal configuration and the traditional arguments invoking the third law of thermodynamics to prevent the formation of a naked singularity cannot be used. To go around this issue, Garfinkle et-al claimed that the flux radiation at infinity goes to zero as the black hole becomes extremal. In such case, a discontinuous jump in the temperature from a finite value to zero takes place. This rather unusual behaviour has been used to test alternative definitions of surface gravity which are more suitable to study similar situations (c.f. Marques \cite{marques}). In particular Hayward \cite{hayward2}, Nielsen-Visser \cite{nielsen} definitions are `well-behaved' at the extremal point.

Leaving aside an attempt to provide an alternative definition of surface gravity for a generic horizon, we analyse the thermodynamics of the solution found by Garfinkle et-al, in both, Einstein and Jordan, frames. As we will see the conformal factor diverges at the extremal configuration, and one should not expect to find the same physical features in both frames in this particular setting\cite{salgado}. However, in the non-extremal regime, the conformal factor is well defined and one would expect that the thermodynamic features should remain invariant.  In particular, the peculiar singularity structure of the the  solution suggests that a holographic argument may be used to provide a suitable expression for the entropy of the connected region of the spacetime. This allows us to find a thermodynamic temperature which complies with the requirements expected from the third law of thermodynamics.

This paper is organised as follows. In section \ref{section1} we present the black hole solution obtained by \cite{gibbons1,gmaeda,ghs,ghserrata} in both Einstein and Jordan frames. We take special care of the extremal configurations, noting that the conformal factor relating these two frames diverges in this case. We found that there is a note worthy misprint in the metric for the Jordan frame in \cite{ghs}. This affects the  extremal configuration and alters the location of the even horizon [c.f. \eqref{JF.horizon}]. We observe that the extremal configuration of the Jordan frame solution cannot be the limit of a sequence of solutions. In section III we  compare  the results obtained from calculating the surface gravity using a comprehensive set of definitions. We argue that the standard definition does not yield a physical result since it depends on the conformal frame one uses. Furthermore, the non-trivial singular structure of the solution suggests the use of a holographic argument to remove the micro-states disconnected from the region of spacetime we analyse. This is presented in section IV. Finally, we note the richer thermodynamics of this holographic entropy and its agreement with the third law of black hole mechanics. 


\section{A charged black hole in string theory}
\label{section1}

Let us revise the charged low energy solution found by Gibbons and Maeda \cite{gmaeda} and independently by Garfinkle, Horowitz and Strominger \cite{ghs}. For later reference, we call this solution GMGHS. We consider the solution in both, Einstein and Jordan, frames individually, thus avoiding any possible ambiguity.

\subsection{Einstein frame solution}
	
The static, magnetically charged  solution in spherical symmetry is (see \cite{ghs} together with the erratum \cite{ghserrata} for details)
	\beq
	\label{sol.efmetric1}
	\d s^2_{\text{E}}  	 = - \left[1 - \frac{2M}{r} \right] \d t^2 + \left[ 1 - \frac{2M}{r}\right]^{-1} \d r^2  + r \left[r - \frac{Q^2 e^{-2\phi_0}}{M} \right] \d \Omega^2,
	\eeq
	\beq
	\label{sol.dilaton}
	e^{-2\phi_{\text{E}}} 	 = e^{-2 \phi_0}\left[1 - \frac{Q^2 e^{-2\phi_0}}{M r} \right]
	\eeq
and
	\beq
	\label{sol.faraday}
	F_{\text{E}} 			 = Q \sin\theta \d\theta \wedge \d \varphi.		
	\eeq
	
Here $F$ is the Faraday tensor, $M$ represents the mass of the black hole, $Q$ is the magnetic charge and $\phi$ and $\phi_0$ corresponds to the dilaton field and its  asymptotic value, respectively. Note that the non-angular part of the solution is completely equivalent to that of the Schwarzschild solution. However, this similarity is only apparent since the radial coordinate does not correspond to the areal radius. 

Note that this solution has diverging Ricci and Weyl curvatures ($R^a_{\ a}$ and $C_{abcd}C^{abcd}$) at
	\beq
	r = 0 \quad  \text{and} \quad r_- = \frac{Q^2 e^{-2\phi_0}}{M}.
	\eeq
 There is also a coordinate singularity at $r=2M$.


In the forthcoming analysis, it is sometimes useful to express the metric \eqref{sol.efmetric1} in areal coordinates. To avoid confusion and introducing unnecessary notation we keep the label $t$ for the time coordinate and use  an $R$ to denote the  aereal coordinate. Thus, using the fact that any static, spherically symmetric metric has the general form \cite{nielsen}
	\beq
	\label{sol.areal}
	\d s^2_{\text{E}} = - \sigma(R)^2 f(R)\d t^2 + f(r)^{-1}\d r^2 + R^2 \d \Omega^2,
	\eeq
we see that, in the case of \eqref{sol.efmetric1} we have
	\beq
	f(R) = \frac{1}{4}\left[1 - \frac{4 M^2 e^{2\phi_0}}{Q^2 + \sqrt{Q^4 + 4 e^{4\phi_0} M^2 R^2}} \right] \left[\frac{Q^4 + 4 e^{4\phi_0} M^2 R^2}{e^{4\phi_0} M^2 R^2} \right]
	\eeq
and
	\beq
	\sigma(R) = \frac{2 M R e^{2\phi_0}}{\sqrt{Q^4 + 4 e^{4\phi_0} M^2 R^2}}.
	\eeq
	
Note that this coordinate system does not cover the entire spacetime and some features of its singular structure cannot be seen directly. These coordinates cover the region of spacetime given by $r \geq r_-$ in \eqref{sol.efmetric1}. In fact the curvature singularity located at $r_-$ has the locus $R =0$ in the new coordinate system. 

The horizons are located at
	\beq
	R_{\pm} = \pm \sqrt{4 M^2 - 2 Q^2 e^{-2 \phi_0}},
	\eeq
where the inner horizon has been pushed outside the domain of the new coordinate system.

 The `extremal' configuration occurs when the horizon completely vanishes, that is when
	\beq
	Q^2 = 2 e^{2\phi_0} M^2,
	\eeq
and the metric becomes 
	\beq
	\d s^2_{\text{E}} = -\frac{\sqrt{M^2 + R^2} - M}{\sqrt{M^2 + R^2} + M}\d t^2 + \left[\frac{R^2}{M^2 + R^2}\right]\left[\frac{\sqrt{M^2 + R^2} + M}{\sqrt{M^2 + R^2} - M} \right] \d R^2 + R^2 \d \Omega^2,
	\eeq
which describes a naked singularity. To the best of our knowledge, this solution has not been presented elsewhere and an independent analysis will come in a forthcoming publication.

It follows from the dilaton solution \eqref{sol.dilaton} that the function  $e^{2 \phi}$ evaluated at the horizon diverges in the extremal limit. This is a crucial point since this function is the conformal factor relating the two frames \cite{ghs}. Therefore, we cannot estate that the dynamics of these two frames are equivalent in the extremal case.

\subsection{Jordan frame solution}

In this frame the metric takes the form
	\beq
	\label{sol.jordanm}
	\d s^2_{\text{J}} = -\frac{1- \frac{2 M e^{\phi_0}}{\rho}}{1-\frac{Q^2 e^{-\phi_0}}{M\rho}}\d \tau^2 + \frac{\d \rho^2}{\left( 1-\frac{2 M e^{\phi_0}}{\rho} \right)\left(1-\frac{Q^2 e^{-\phi_0}}{M\rho} \right)} + \rho^2 \d \Omega^2.
	\eeq
This solution can also be found through a conformal transformation of the metric \eqref{sol.efmetric1} followed by a rescaling of the non-angular coordinates given by $\tau = e^{\phi_0} t$ and $\rho = e^{\phi_0} r$, where the temporal and spatial coordinates are labelled by $\tau$ and $\rho$, respectively.

The solution for the electromagnetic field is given by
	\beq
	\tilde{F}_{\text{J}} = \left[e^{2\phi_0} Q \sin \theta\right]\left[1 - \frac{Q^2 e^{-\phi_0}}{M \rho}\right]^{-1}\ \d\theta \wedge \d\varphi.
	\eeq
while for the dilaton field it becomes a formal  expression involving integrals of Heun confluent functions.

It is straightforward to see that the metric for the extremal configuration is
	\beq
	\label{jordanext}
	\d s^2_{\text{J}} = - \d \tau^2 + \left[ 1 - \frac{2 M e^{\phi_0}}{\rho}\right]^{-2} \d \rho^2 + \rho^2 \d \Omega^2.
	\eeq
Notice that in the original work of Garfinkle et-al \cite{ghs} there is a misprint in the exponential factor accompanying $Q^2$ (note that this does not correspond to the erratum \cite{ghserrata}). In their work it appears to be  $e^{3\phi_0}$ where we have $e^{-\phi_0}$. This seemingly harmless change is important since the extremal limit changes drastically, as one can verify directly if one uses their expression. However, the extremal metric, equation (15) in \cite{ghs}, is the correct  limit of \eqref{sol.jordanm}, above. 

The horizon is located at
	\beq
	\label{JF.horizon}
	\rho_+ = 2 M e^{\phi_0},
	\eeq
and quick inspection of the Ricci and Weyl curvature scalars reveals that, in addition to the singularity at the origin, the curvature diverges at
	\beq
	\rho_- = \frac{Q^2 e^{-\phi_0}}{M}.
	\eeq
This separates the spacetime into two disconnected regions divided by a singular shell. Moreover, in this case $\rho$ is an  areal coordinate and the event horizon at $\rho = \rho_+$ depends only upon the mass of the black hole. Therefore, if we make Hawking's identification of entropy with a quarter of the area of the event horizon, and take the mass and charge as the other two thermodynamic degrees of freedom, we would not obtain a consistent first law, as we will see in the next section.  

\begin{figure}
	\includegraphics[width=.5\columnwidth]{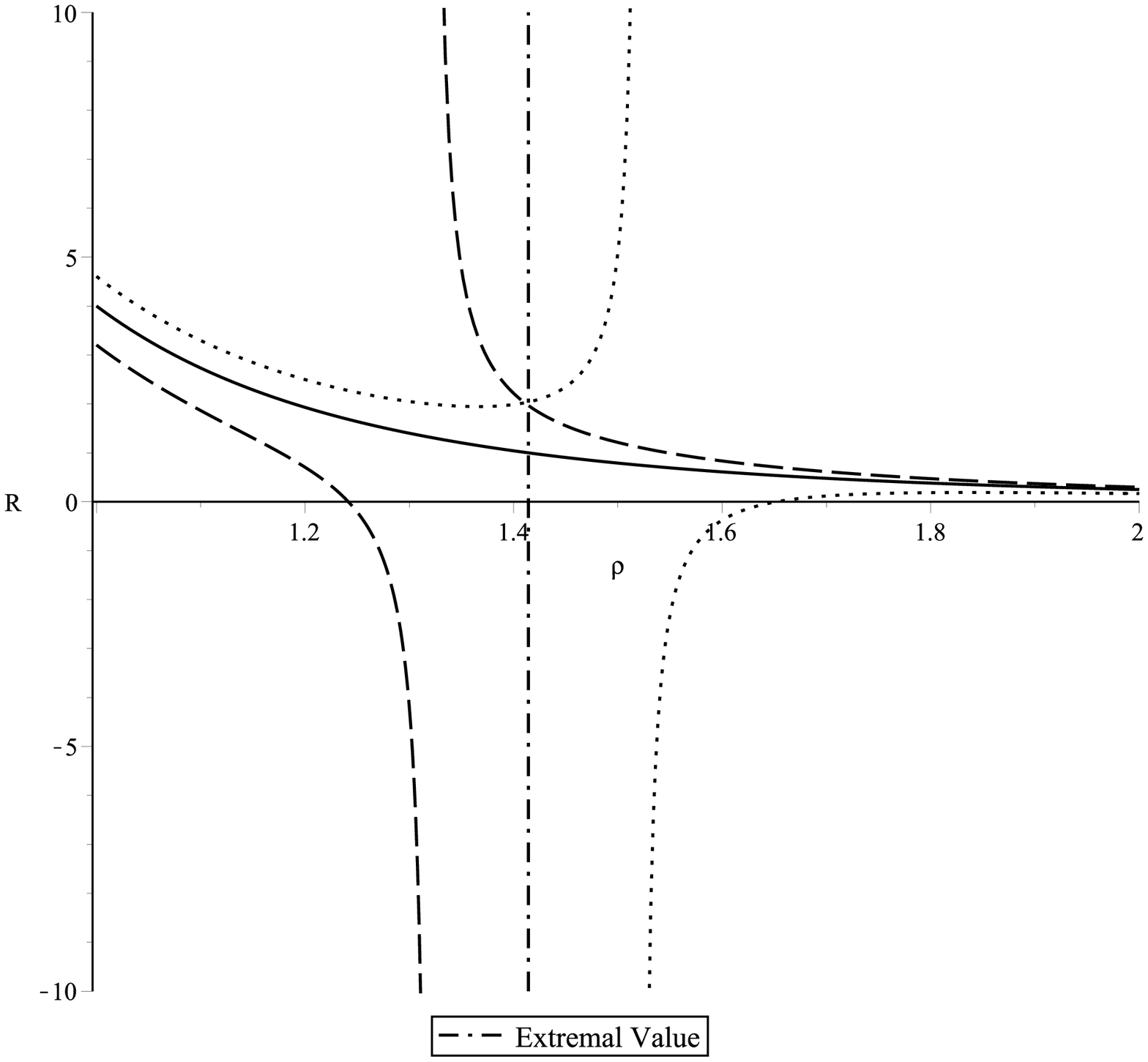}\includegraphics[width=.5\columnwidth]{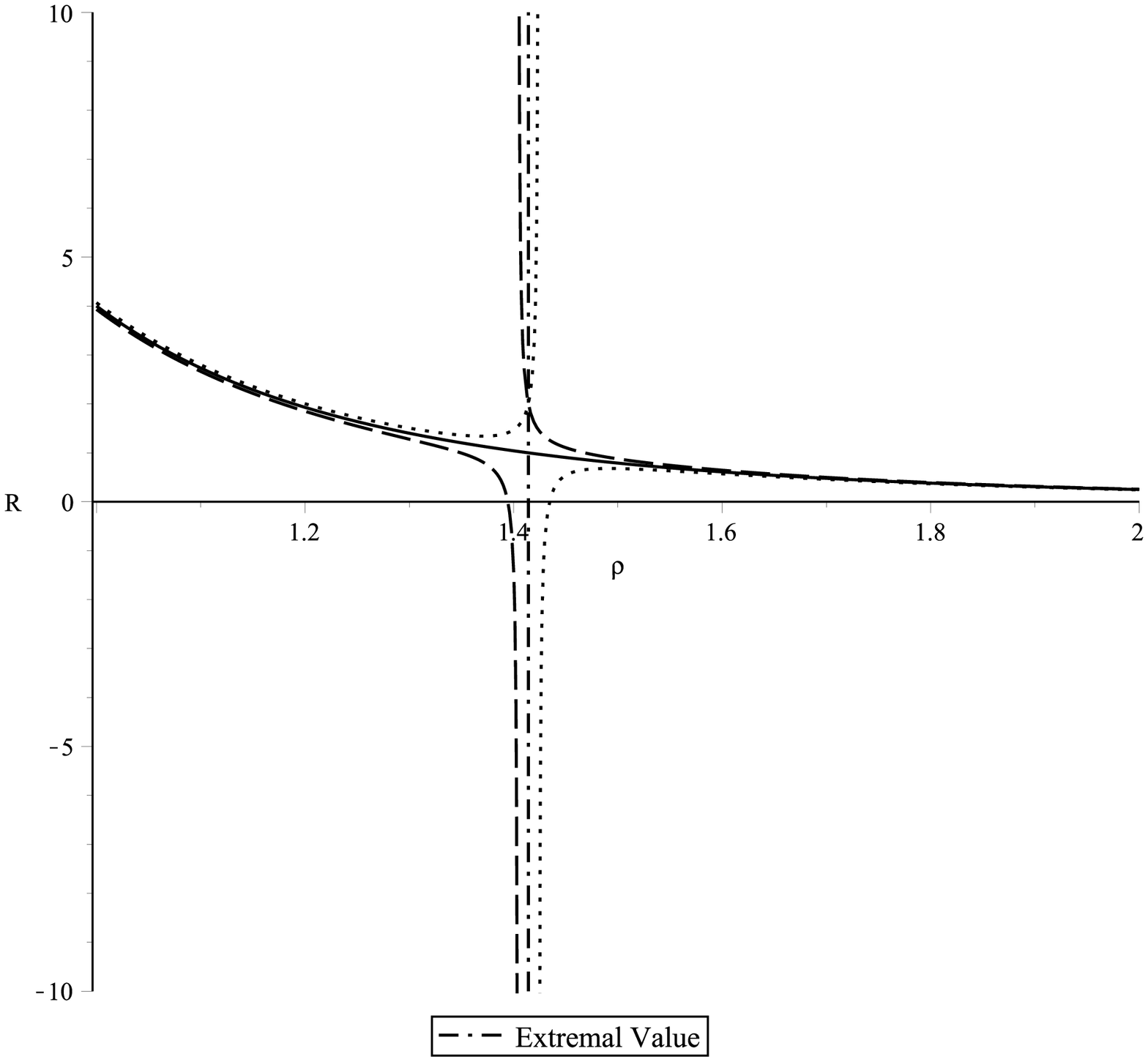}
	\caption{This figure shows the curvature scalar of a sequence of neighbouring solutions to the extremal configuration of the GMGHS black hole. The continuous line represents the curvature of the extremal solution, whereas the dotted lines are those for which $M = M_{\text{extremal}} \pm \epsilon$. These lines diverge at the value $\rho=\rho_-$, except for the extremal one. Here, we are using $Q=1$ and $\phi_0 = 0$ for the charge and dilaton, respectively. We take $\epsilon = 0.05, 0.005$ from left to right. }
	\label{fig.curvature}
	\end{figure}

The extremal configuration occurs when
	\beq  \label{extremal}
	Q^2 = 2 e^{2\phi_0} M^2,
	\eeq
and has the feature of removing the shell singularity from the solution. To see this, let us explore the curvature scalar of neighbouring solutions of the extremal configuration \eqref{jordanext}.  There is no longer an event horizon and all that is left is a naked singularity at the origin. In figure \ref{fig.curvature} we show a sequence of plots for the curvature scalar of various neighbouring solutions of the extremal configuration. We observe that the extremal metric is not a cluster point in the space of solutions. Moreover, no perturbation of the extremal configuration reproduces the singularity structure of the Jordan frame metric, equation \eqref{sol.jordanm}.

From the above discussion it is clear that we are dealing with two solutions which fail to be conformally related in the extremal case. In the coming sections, we make a thermodynamic analysis for each solution separately.


\section{Entropy and surface gravity of the GMGHS black hole}

In the traditional physics jargon, surface gravity has become a synonym of temperature. Indeed, in stationary configurations, the surface gravity is an invariant quantity unambiguously defined once a normalisation of timelike Killing vector is set.   In this section we recall a set of definitions of surface gravity obtained by various authors and compare their near extremal behaviour in the GMGHS solution. We proceed as in the previous section and make a separate thermodynamic analysis in each frame. Here, we limit ourselves to explore how  different definitions of surface gravity match the empirical behaviour one would expect from the laws of black hole mechanics. 

The definitions of surface gravity that we use are all valid for spherical symmetry. Our selection is by no means exhaustive but we have chosen a few representatives to illustrate the different physical predictions arising from a seemingly harmless ambiguity, which we explain below.

The standard procedure to find the surface gravity of a stationary spacetime is through its timelike Killing vector filed, $\xi^a$.  At the horizon, this vector field becomes null and the surface gravity is  defined as the proportionality factor measuring  its departure from a geodesic curve,
	\beq
	\xi^b \nabla_{b} \xi^{a} = \kappa \xi^a.
	\eeq 
It is conventional to adopt a unit normalisation at infinity in order to remove the ambiguity arising  from rescaling of the null Killing vector. Note that this is valid only in the case of asymptotically flat spacetimes. In such case, the surface gravity is simply given by \cite{wald}
	\beq
	\label{kappa1}
	\kappa_{1} = \left[-\frac{1}{2} \nabla_a \xi_b \nabla^a \xi^b\right]^{1/2}.
	\eeq

In the case of spherical symmetry there is a number of ways in which one can calculate the surface gravity. If the spacetime is also static, all of these are equivalent. Indeed this is true for Schwarzschild and Reissner-Nordstr\"om black holes but not in general, as we show below.
    
Let us begin with one of the most widely used definitions of surface gravity for a static and spherically symmetric spacetime \cite{wald}
	\beq
	\kappa_{2} = \lim_{r\rightarrow r_\text{h}} \left[\frac{1}{2\sqrt{-g_{tt}g_{rr}}}\frac{\partial g_{tt}}{\partial r} \right].
	\eeq

In the case of a dilaton black hole (dirty black hole), Visser has shown that an equivalent way of calculating the surface gravity is given by \cite{visser}
	\beq
	\kappa_{3} = \frac{1}{2 r_{\text{h}}}e^{-\phi(r_{\text{h}})}\left[1 - b'(r_{\text{h}}) \right],
	\eeq
where the function $b(r)$ is obtained by expressing the time and radial components of the metric in the form
	\beq
	g_{tt} = e^{-2\phi(r)}\left[1 - \frac{b(r)}{r} \right] \quad \text{and} \quad g_{rr} = \left[1- \frac{b(r)}{r} \right]^{-1},
	\eeq
and where the prime denotes differentiation with respect to  $r$. 

These definitions are completely equivalent in each frame and give the values
	\beq
	\kappa^{\rm  E} = \frac{1}{4 M} \quad \text{and} \quad \kappa^{\rm J} = \frac{1}{4 M e^{\phi_0}},
	\eeq
in the Einstein and Jordan frames, respectively. Indeed, these are frame independent only for the case where the conformal factor at infinity is the identity, i.e. when $\phi_0 = 0$ (c.f. Jacoboson \cite{jacobson01}).

There is a more general  definition introduced by Hayward \cite{hayward2} valid also in the dynamical (non-static), but still spherically symmetric, scenario .  It is obtained by considering the geometry of the two-surface formed by the time and radial coordinates of a spherically symmetric spacetime, and it is given by
	\beq
	\kappa_{4} = \frac{1}{2} \star_2 \d \star_2 \d r,
	\eeq
where $\star_2$ represents the Hodge dual operator of the two dimensional space normal to the spheres of symmetry. In this case we have
	\beq
	\kappa^{\rm E}_4 = \frac{1}{8}\left[4 M^2 e^{2\phi_0} - Q^2 \right]\left[M^2 e^{\phi_0}\sqrt{4 M^2 e^{2\phi_0} - 2Q^2} \right]^{-1},
	\eeq
in the Einstein frame, whereas
	\beq
	\kappa^{\rm J}_4 = \frac{1}{8}\left[2 M^2 e^{2 \phi_0} - Q^2\right]\left[M^3 e^{3\phi_0}\right]^{-1}
	\eeq
in the Jordan frame. In this case, the surface gravity is not conformally invariant regardless of the asymptotic value of the dilaton. Any prescription that gives different values in the two frames is clearly not the right definition to use.

Alternatively, one can use Euclidean techniques to find the surface gravity. In this case, one starts by transforming the metric into Kruskal-Szekeres type coordinates and identifying the period of the Euclidean section in the last step of the coordinate transformation. This elegant approach, however, is not free of ambiguities either  \cite{nielsen3} . Let us write the final coordinate relation as
	\beq
	\label{sg.euclid}
	{\rm i}\bar t 	 = F(r) \sin\left(\frac{1}{2 r_\text{H}} t\right) \quad \text{and} \quad%
	\bar x 			 = F(r) \cos\left(\frac{1}{2 r_\text{H}} t\right)
	\eeq
where $r_\text{H}$ stands for the aereal value of the horizon $R_+$ and $\rho_+$, respectively.  The Euclidean section radial functions are
	\begin{align}
	F_\text{E}(r)		= &\left(\cosh\alpha_\text{E} + \sinh\alpha_\text{E} \right)\left[Q^2 e^{-2\phi_0} + \sqrt{4 M^2 R^2 + Q^4 e^{-4\phi_0}} - 4 M^2 \right]^{\frac{1}{2}\frac{M}{R_+}},\\
	F_\text{J}(\rho)	= &\left(\cosh\alpha_\text{J} + \sinh\alpha_\text{J} \right) \sqrt{\rho - 2 M e^{\phi_0}},
	\end{align}
with
	\beq
	\alpha_\text{E} 	 = \frac{e^{\phi_0}}{4 M}\sqrt{\frac{4 M^2 R^2 + Q^4 e^{-4\phi_0}}{4M^2 e^{2\phi_0} - 2 Q^2}} \quad \text{and} \quad
	\alpha_\text{J}		 = \frac{1}{4}\frac{\rho}{M e^{\phi_0}}
	\eeq
for the Einstein and Jordan frames, respectively. We can read out the period $\beta$ of the Euclidean section directly from \eqref{sg.euclid}. This will be proportional to the temperature of the black hole and  we can define the Euclidean surface gravity as 
	\beq
 	\kappa_{5} = \frac{2 \pi}{\beta}.
	\eeq
These are
	\beq
	\kappa^{\rm E}_5 = \frac{1}{2}e^{\phi_0} \left[\sqrt{4 M^2 e^{2\phi_0} - 2 Q^2} \right]^{-1} \quad \text{and} \quad \kappa^{\rm J}_5 = \left[4 M e^{\phi_0}\right]^{-1}.
 	\eeq
	
For comparison, let us consider also the surface gravity of the Reissner-Nordstr\"om black hole \cite{deFelice,HawkingEllis}. Here, one would have
	\beq
	\kappa^{\rm E}_{6}  = \frac{1}{2 R_+} \quad \text{and} \quad  \kappa^{\rm J}_{6} = \frac{1}{2}\frac{\rho_+ - \rho_-}{ \rho_+^2}.
	\eeq
Interestingly, for the GMGHS solution in the Einstein frame the value of $\kappa_6$ coincides with that of the Euclidean surface gravity, $\kappa_5$, while in the Jordan frame it is equal to the dynamic surface gravity $\kappa_4$. Note that here $\rho_-$ is a curvature singularity, as oppossed to the Reissner-Nordstrom case. This excercise does not intend to provide a definition for the surface gravity, but merely as reference point.

One can also use the first law of thermodynamics to find the temperature, and hence the surface gravity. To do this, we must have an expression for the entropy and use it as the thermodynamic fundamental relation, whose degrees of freedom are $M$ and $Q$. Thus, the `thermodynamic' surface gravity is given by
	\beq
	\label{sg.gibbs}
	\kappa_{7}  = 2 \pi \left[\frac{\partial S}{\partial M} \right]^{-1}.
	\eeq

It is important to note that this last expression is well defined  only when a true thermodynamic fundamental relation is provided. That is, when the thermodynamic potential depends \emph{at least} on two degrees of freedom. The contact structure of thermodynamics demands that $dM$ in the first law to be an exact form given the thermodynamic fundamental relation $M$. If we were to write the first law of thermodynamics of a one degree of freedom system $dM = T(S)dS$, that requirement would not be satisfied, hence formally that does not qualify as a thermodynamic system. Nevertheless, one can find this situation as a limit of a true thermodynamic system, e.g. Schwarzschild black hole as the limit of vanishing charge in the Reissner-Nordstr\"on case. In this case, if we take the mass as our thermodynamic potential, we must have $M = M(S,Q)$ or, equivalently, $S = S(M,Q)$.  In the Einstein frame the entropy is given by 
	\beq
	\label{sg.entropy1}
	S_{\text{E}} = \frac{1}{4} A_{\text{E}} = \left[4 M^2 - 2 Q^2e^{-2\phi_0}\right]\pi.
	\eeq
 However, when the same argument is applied in the Jordan frame we obtain
	\beq
	\label{sg.entropyJ}
	S_{\text{J}} = \frac{1}{4}A_{\text{J}} = \pi \rho_+^2 = 4 \pi M^2 e^{2\phi_0}.
	\eeq
This is not a good fundamental relation and we cannot write down a Gibbs one-form for the first law of thermodynamics in the entropy representation. Nevertheless, the thermodynamic surface gravity can be easily calculated, although we cannot expect it would be of any physical relevance.  Also, note that the fundamental relation, equation \eqref{sg.entropy1}, is a separable function, and thus the temperature does not depend on $Q$. The corresponding surface gravities are
	\beq
	\kappa_7^{\rm E} = \frac{1}{4 M}\quad \text{and} \quad \kappa^{\rm J}_7 = \frac{1}{4 M e^{2 \phi_0}}.
	\eeq

\section{Frame independent thermodynamics}

In the above discussion, we have tested a number of well known definitions for the surface gravity of static and spherically symmetric spacetimes and obtained results which are clearly not frame independent. There is a general framework for finding the black hole entropy, and therefore its temperature, in any diffeomorphism invariant theory of gravity, namely the Noether charge method of Wald \cite{waldNQ}
	\beq
	\label{waldentropy}
	S_W = 2 \pi \oint_\Sigma {\bf Q} 
	\eeq
where the Noether charge ${\bf Q}$ depends on the specific form of the gravity sector of Lagrangian for the Hilbert action and $\Sigma$ is the (two-dimensional) `interior boundary' of an asymptotically flat hypersurface. In our case, these are only functions of the curvature scalar, i.e. $-R$ and $-e^{-2 \phi} R$ for the Einstein and Jordan frames, respectively and $\Sigma$ corresponds to the horizon. Thus, equation \eqref{waldentropy} is simply written as  \cite{jacobson01,visser02}
	\beq
	S_W^{\rm E} = \frac{1}{4} \int_{H} \sqrt{\ _2 g} \d^2 x \quad \text{and} \quad S^{\rm J}_W = \frac{1}{4} \int_{\tilde H}e^{-2\phi(\tilde H)} \sqrt{\ _2 {\tilde g}} \d^2 {\tilde x}.
	\eeq
Here, we have denoted by $\ _2 g$ the induced metric on the horizon two-surface and, as before, we use a tilde to express the quantities in the Jordan frame. Thus, we obtain the frame independent entropy
	\beq
	S_W = \left(4 M^2 - 2 Q^2 e^{-2 \phi_0}\right) \pi.
	\eeq
As expected, this entropy coincides with the Bekenstein entropy in the Einstein frame. In the Jordan frame, in spite of not reproducing the one quarter of the area entropy, Wald's entropy provides a true thermodynamic fundamental relation.  Another nice feature of Wald's entropy is that it smoothly vanishes as the extremal configuration \eqref{extremal} is approached. The associated temperature $(\partial S/\partial M)^-1$ is given by
	\beq
	T_W = \frac{1}{8 \pi M}.
	\eeq
From this, we obtain the invariant surface gravity 
	\beq
	\kappa_{\rm W} = \frac{1}{4 M}
	\eeq
for both frames. Note that, albeit independent of the conformal frame, Wald's temperature for this particular black hole does not agree with the statement of the third law of black hole mechanics \cite{bardeen}.

\begin{figure}
		\includegraphics[width=.75\columnwidth]{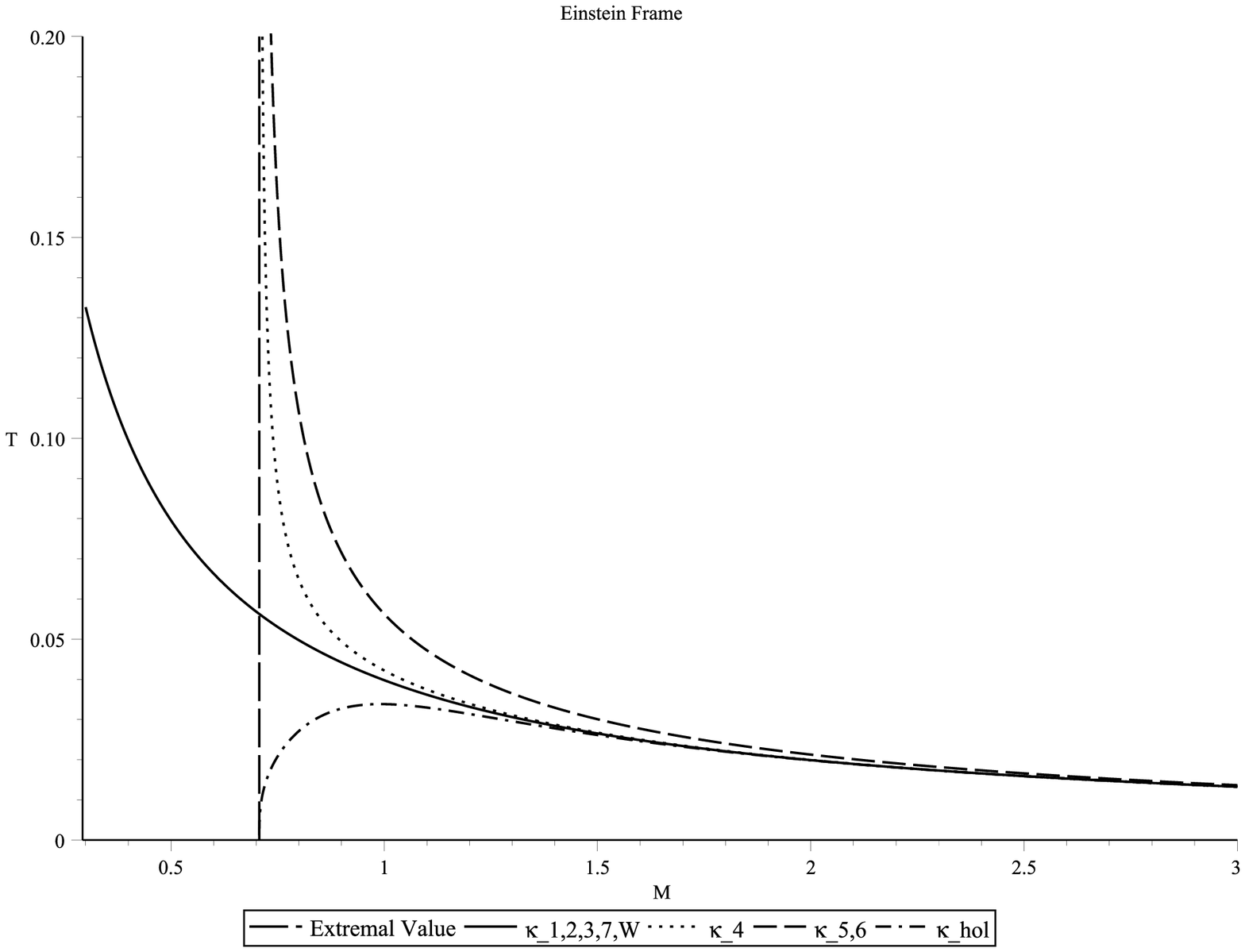}\\
		\includegraphics[width=.75\columnwidth]{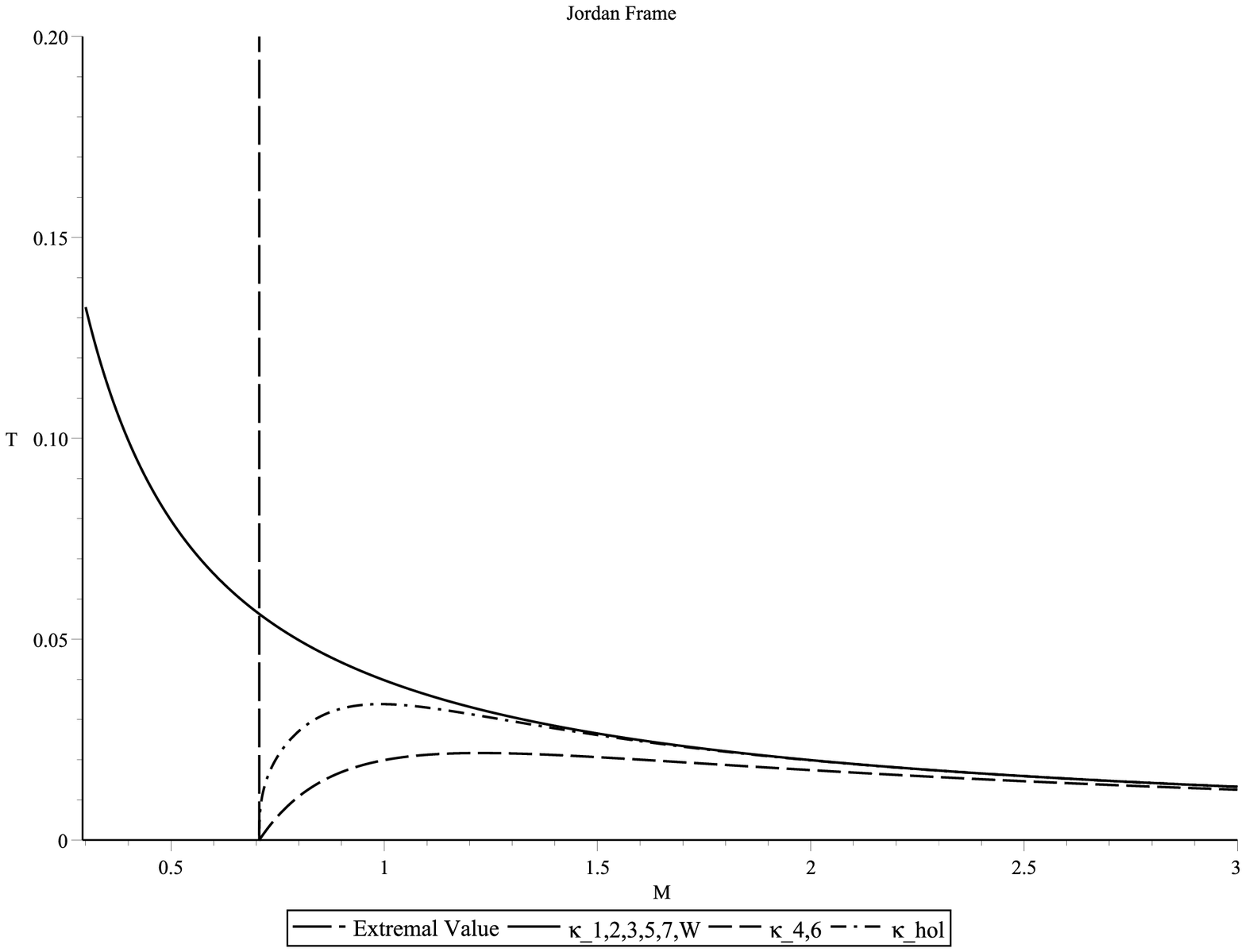}
		\caption{Temperatures associated with the surface gravity definitions presented in table \ref{table}. Here we have fixed the value of the charge $Q=1$ and the value $\phi_0 = 0$. Thus, we see that the solid line corresponds to those temperatures which are conformally invariant. Note that $\kappa_{\rm hol}$ is also invariant and vanishes identically when the extremal configuration is reached, in agreement with the third law of thermodynamics.}
		\label{fig.temps}
	\end{figure}

Now we introduce a general definition of surface gravity for space-times whose singularity structure exhibit two disconnected regions. Suppose that there are two singularities both surrounded by an event horizon located at $f_{\rm{hor}}(x^a) = 0$, where $x^a$ are coordinates for the space-time. The interior singular surface is described by the locus $f_{\rm{int}}(x^a) = 0$ and the exterior one by $f_{\rm{ext}}(x^a) = 0$. All the volume enclosed by the exterior singularity cannot be physically connected to the region outside of it. Therefore, in a semi-classical setting,  one might be over-counting the internal micro-states in the expression for the entropy
        \beq
        \label{gen.entropy}
        S = \frac{1}{4} A_{\rm{H}},
        \eeq
where $A_{\rm{H}}$ is the area of the  event horizon surrounding both singularities. Whenever the interior and exterior singularities coincide in a single point located at the origin of coordinates we return to this usual expression for the entropy \eqref{gen.entropy}. If we want to make an appropriate use of the holographic principle we need to remove those microscopic degrees of freedom which cannot have an imprint in the event horizon. The simplest way of doing this, is by removing the region enclosed by the surface $f_{\rm{ext}}(x^a) = 0$, and define an effective area $A_{\rm{hol}}$ enclosing the volume of the region between the exterior singularity and the exterior event horizon. Thus, the entropy will be  given by
        \beq
        \label{eff.entropy}
        S_{\rm{hol}} = \frac{1}{4} A_{\rm{hol}}.
        \eeq
As the minimal area enclosing a given volume is a sphere, we can define an effective radius $\varrho_{\rm{eff}}(x^a)$. Hence the entropy relation can be expressed as
        \beq
        \label{eff.entropyradius}
        S_{\rm{hol}} = \pi \varrho_{\rm{eff}}^2.
        \eeq
This is particularly simple to accomplish for spherically symmetric spaces as it is the GMGHS solution in both frames. That is, removing the region enclosed by the exterior singular surface of radius $\varrho_-$ centered at the origin, and defining the effective area enclosing the volume of the region $\varrho_- < \varrho < \varrho_+$
        \beq
        A_{\rm{hol}} = 4\pi \varrho_{\text{eff}}^2= 4 \pi \left(\varrho_+^3 - \varrho_-^3 \right)^\frac{2}{3}, 
	\eeq  
we arrive at the holographic entropy 
	\beq  \label{holent}
	 S_{\rm{hol}} = \frac{A}{4} =  \pi \left(\varrho_+^3 - \varrho_-^3 \right)^\frac{2}{3}.
	\eeq
It is remarkable that such an entropy obtained by a holographic inspired argument is frame independent and, in this  particular case, takes the form
	\beq
        S_{\rm{hol}} = \pi e^{2\phi_0} \frac{\left[8 M^6 - Q^6 e^{-6\phi_0}\right]^{\frac{2}{3}}}{M^2}.
	\eeq
This expression also vanishes in the extremal limit and its associated temperature,
	\beq
	T_{\rm hol} = \frac{M}{2 \pi}  \frac{\left[\varrho_+^3 - \varrho_-^3 \right]^\frac{1}{3}}{\varrho_+^3 + \varrho_-^3},
	\eeq
complies with the requirement of the third law of black hole mechanics from which we read off the surface gravity [c.f. figure \ref{fig.temps}]
	\beq
	\kappa_{\rm hol} = M \frac{\left[\varrho_+^3 - \varrho_-^3 \right]^\frac{1}{3}}{\varrho_+^3 + \varrho_-^3}.
	\eeq


The outcome of the different definitions presented above is summarised in table \ref{table}. 
	
	\begin{table}
	\begin{tabular}{|c|c|c|}
	\hline
			& {\bf Einstein frame}	& {\bf Jordan frame}	\\
	\hline
	$\kappa_{1,2,3}$ 		&  $\left[4 M \right]^{-1}$		& $\left[4 M e^{\phi_0}\right]^{-1}$		\\[0.1cm]
	\hline
	$\kappa_{4}$		& $\frac{1}{8}\left[4 M^2 e^{2\phi_0} - Q^2 \right]\left[M^2 e^{\phi_0}\sqrt{4 M^2 e^{2\phi_0} - 2Q^2} \right]^{-1}$					&  $\frac{1}{8}\left[2 M^2 e^{2 \phi_0} - Q^2\right]\left[M^3 e^{3\phi_0}\right]^{-1}$ 				\\[0.1cm]
	\hline
	$\kappa_{5}$		&	$\frac{1}{2}e^{\phi_0} \left[\sqrt{4 M^2 e^{2\phi_0} - 2 Q^2} \right]^{-1}$				&  $\left[4 M e^{\phi_0}\right]^{-1}$				\\[0.1cm]
	\hline
	$\kappa_{6}$		&	$\frac{1}{2}e^{\phi_0} \left[\sqrt{4 M^2 e^{2\phi_0} - 2 Q^2} \right]^{-1}$				&  $\frac{1}{8}\left[2 M^2 e^{2 \phi_0} - Q^2 \right]\left[M^3 e^{3\phi_0}\right]^{-1}$				\\[0.1cm]	
	\hline
	$\kappa_{7}$		&	$\left[4 M \right]^{-1}$				&	$\left[4 M e^{2\phi_0} \right]^{-1}$			\\[.1cm]	
	\hline
	$\kappa_{\rm W}$	&					$[4 M]^{-1}$						& 	$[4 M]^{-1}$\\[.1cm]
	\hline
	$\kappa_{\rm hol}$	& $M^3 e^{2\phi_0}\left[8 M^6 e^{6 \phi_0} - Q^6  \right]^{\frac{1}{3}} \left[8 M^6  e^{6\phi_0}+ Q^6  \right]^{-1}$ & $M^3 e^{2\phi_0}\left[8 M^6 e^{6 \phi_0} - Q^6  \right]^{\frac{1}{3}} \left[8 M^6  e^{6\phi_0}+ Q^6  \right]^{-1}$\\[.1cm]
	\hline
	\end{tabular}
	\caption{This table shows the different results obtained after calculating the surface gravity in each frame. We observe that: (a) $\kappa_{1,2,3}$ and $\kappa_7$ are frame invariant only when the asymptotic value of the dilaton vanishes. (b) Hayward's definition, $\kappa_4$, shows a more interesting behaviour in the near extremal regime. However, it is not possible to make it frame independent. (c) There is an interesting comparisson between the unrelated $\kappa_4$ and $\kappa_6$ (RN) in the Jordan frame. (d) The only two frame independent definitions, regardless of the asymptotic value of the dilaton, are $\kappa_{\rm W}$ and $\kappa_{\rm hol}$.}
	\label{table}
	\end{table}


	\begin{figure}
		\includegraphics[width=.75\columnwidth]{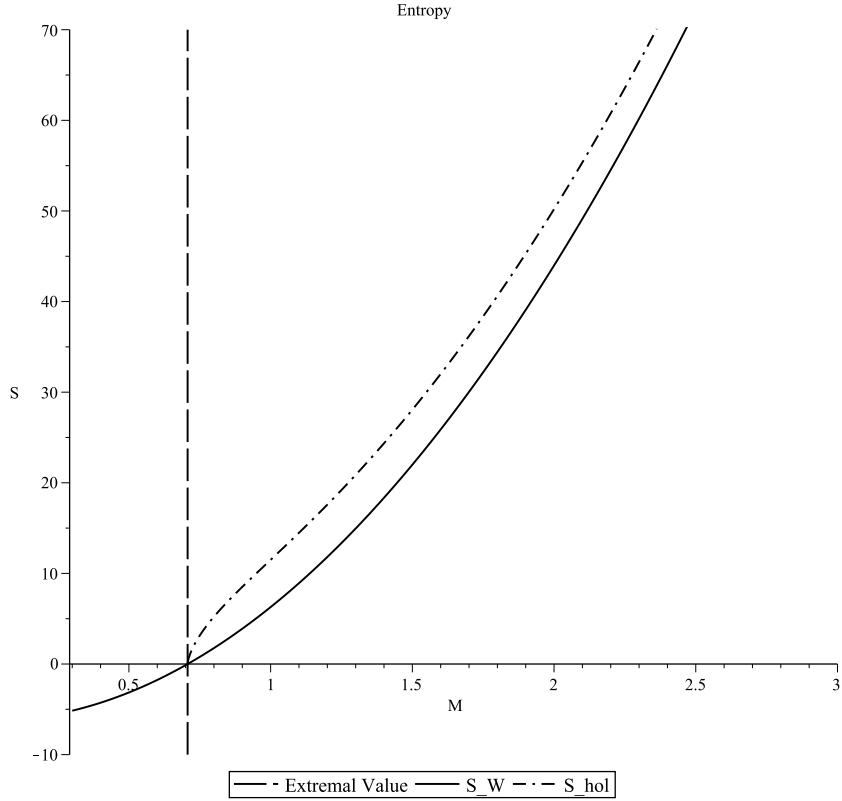}\\
		\caption{The behaviour of the two invariant entropies near the extremal configuration.}
		\label{fig.ents}
	\end{figure}

	\begin{figure}
		\includegraphics[width=.75\columnwidth]{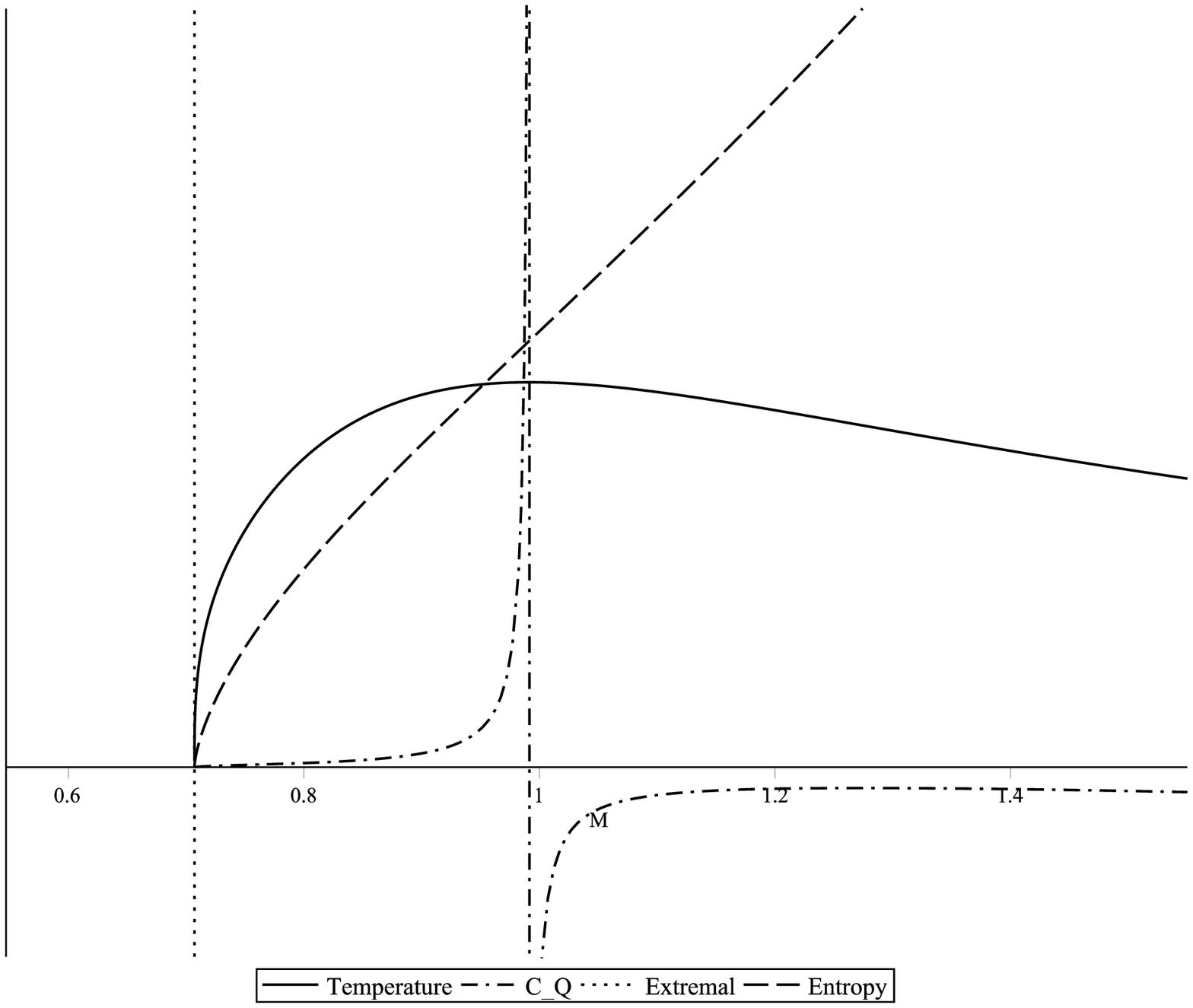}\\
		\caption{This figure shows the phase transition point sinalised by a divergent behaviour of the heat capacity $C_Q$. Note that this point corresponds to the global maximum of the temperature $T_{\rm hol}$ and to the change of convexity of the entropy $S_{\rm hol}$.}
		\label{fig.tent}
	\end{figure}

\section{Closing remarks}

In this paper we have analysed the low energy solution found independently by Gibbons and Maeda and Garfinkle, Horowtiz and Strominger. We verified that the conformal factor between the Einstein and Jordan frames is not well defined in the extremal limit and, therefore, a different conformal transformation is requiered to relate the extremal black hole in the Einstein and Jordan frames. This was a crucial point in our discussion since we were interested in  the near extremal behaviour of the solution. Moreover, by inspecting the curvature scalar of the Jordan frame solution, we observed that the extremal metric cannot be a cluster point in the space of solutions to the Einstein field equations. Therefore, our analysis of this particular solution was intended to explore the invariance of various entropy and surface gravity definitions under a change of conformal frame and to observe if their behaviour is in agreement with the third law of black hole mechanics.

Motivated by a recent study of the thermodynamic geometry of this solution  in the vicinity of the extremal configuration  \cite{comment}, we look at a comprehensive  - although not exhaustive - set of definitions for the surface gravity. We observe that $\kappa_1$, $\kappa_2$, $\kappa_3$ and $\kappa_7$ are invariant only if the conformal factor relating the two frames is normalized to unity at infinity, which coincides with what is estated in \cite{jacobson01}. In the case of $\kappa_4$, $\kappa_5$ and $\kappa_6$ we see that they are not frame invariant independently of the normalization of the conformal factor (as recently discussed in \cite{nielsen_arxiv}). Finally, we studied Wald's entropy definition of entropy as a Noether charge together and an entropy inspired by the holographic principle given the singular structure of the solution. We found that both definitions are frame invariant and consequently both give an invariant surface gravity. However, these last two entropies yield different thermodynamic descriptions. On the one hand, Wald's entropy vanishes as the extremal configuration is approached, but its associated  temperature remains finite and independent of the charge and the asymptotic value of the dilaton, failing to behave correctly according to the third law of black hole mechanics. On the other hand,  the propossed holographic definition for the entropy, together with its temperature, vanishes when the black hole becomes extremal. It is also interesting that this entropy presents a phase transition signalised by the divergence of the heat capacity at constant charge $C_Q$ [c.f. figure \ref{fig.tent}]. Furthermore, this phase transition corresponds to the maximum temperature of the black hole and, therefore, the heat capacity changes sign, indicating the transition from a stable to an unstable thermodynamic configuration.

The invariant nature of the holographic entropy presented here, dependes solely on the singularity structure of the solution. This suggests that $S_{\rm hol}$ has a topological nature. This idea will be explored in the future.

 In sum, the equivalence of thermodynamic quantities in two conformally related theories is a delicate issue. In this case,  there is no clear argument to prefer a particular definition other than its invariance properties. In this work, we took the third law of black hole mechanics as an additional guide to choose a more suitable definition for the entropy of a GMGHS black hole.

\section*{Acknowledgements}

FN receives support from DGAPA-UNAM (postdoctoral fellowship). CSLM would like to thank the Wigner Institute for their kind hospitality during the writing of this manuscript. This work was supported by CONACYT project No. 166391 and DGAPA-UNAM No. IN106110.


%

\end{document}